\documentclass[aps,pra,twocolumn,showpacs,floatfix]{revtex4}
\usepackage{graphicx}
\usepackage{amsfonts,amsmath}
\usepackage{bbm}
\usepackage{color}
\newcommand{\ket}[1]{\vert #1 \rangle}

\newcommand{\ketbra}[2]{\vert #1 \rangle \langle #2 \vert}
\newcommand{\imm}{{\rm i }}

\newcommand{\stso}{{\hbox{STS}_1}}
\newcommand{\stst}{{\hbox{STS}_2}}
\newcommand{\tr}{\hbox{Tr}}
\newcommand{\Id}{{\mathbbm I}}
\newcommand\REVISION[1]{{#1}}
\begin{document}
\title{About the use of fidelity to assess quantum resources}
\author{Matteo Bina}
\affiliation{Dipartimento di Fisica dell'Universit\`a degli 
Studi di Milano, 20133 Milano, Italia}
\author{Antonio Mandarino}
\affiliation{Dipartimento di Fisica dell'Universit\`a degli 
Studi di Milano, 20133 Milano, Italia}
\author{Stefano Olivares}
\affiliation{Dipartimento di Fisica dell'Universit\`a degli 
Studi di Milano, 20133 Milano, Italia} 
\affiliation{CNISM, UdR Milano Statale, I-20133 Milano, Italy.}
\author{Matteo G.~A.~Paris}
\email{matteo.paris@fisica.unimi.it}
\affiliation{Dipartimento di Fisica dell'Universit\`a degli 
Studi di Milano, 20133 Milano, Italia}
\affiliation{CNISM, UdR Milano Statale, I-20133 Milano, Italy.}
\date{\today}
\begin{abstract}
Fidelity is a figure of merit widely employed in quantum technology in
order to quantify similarity between quantum states and, in turn, to
assess quantum resources or reconstruction techniques. Fidelities higher
than, say, 0.9 or 0.99, are usually considered as a piece of evidence to
say that two states are very close in the Hilbert space. On the other
hand, on the basis of several examples for qubits and continuous
variable systems, we show that such high fidelities may be achieved by
pairs of states with considerably different physical properties,
including separable and entangled states or classical and nonclassical
ones. We conclude that fidelity as a tool to assess quantum resources
should be employed with caution, possibly combined with additional
constraints restricting the pool of achievable states, or only as a mere
summary of a full tomographic reconstruction.
\end{abstract}
\pacs{03.67.-a, 42.50.Dv, 03.67.Mn}
\maketitle
\section{Introduction}
In the last two decades several quantum-enhanced communication 
protocols and measurement schemes have been suggested and demonstrated. 
The effective implementation of these schemes crucially relies on the 
generation and characterization of nonclassical states and operations
(including measurements), which represent the two pillars of quantum technology.
The assessment of quantum resources amounts to make quantitative statements 
about the similarity of a quantum state to a target one, or to measure 
the effectiveness of a reconstruction technique. For these purposes one 
needs a figure of merit to compare quantum states. Among the possible 
distance-like quantities that can be defined in the Hilbert space a widely 
adopted measure of closeness of two quantum states is the \textit{Uhlmann 
Fidelity} \cite{Uhl} defined as
\begin{equation}\label{fidelity}
F(\rho_1,\rho_2)= \left( \tr \sqrt{\sqrt{\rho_1} 
\rho_2 \sqrt{\rho_1} }  \right)^2
\end{equation} 
which is linked to the Bures distance
$D_B(\rho_1,\rho_2)=\sqrt{2[1-\sqrt{F}]}$
between the two states $\rho_1$ and $\rho_2$, and provides bounds to 
the trace distance 
\cite{fuc99}
$$1-\sqrt{F(\rho_1,\rho_2)}\leq \frac12 || \rho_1-\rho_2||_1\leq
\sqrt{1-F(\rho_1,\rho_2)}\,.$$
Fidelity is bounded to the interval $[0, 1]$, and values 
above a given threshold close to unit, say, 0.9 or 0.99 are 
usually considered very high. Indeed, this implies that the two
states are very close in the Hilbert space, as it follows from the above
relations between the fidelity and the Bures and trace distances. On the other hand,
neighboring states may not share nearly identical physical properties
\cite{edvd,dodonov}
as one may be tempted to conclude. The main purpose of this paper is to show,
on the basis of several examples for qubits and continuous variable (CV)
systems, that very high values of fidelity may be
achieved by pairs of states with considerably different physical properties,
including separable and entangled states or classical and nonclassical ones.
Furthermore, we provide a quantitative analysis of this discrepancy.
\par
\begin{figure}[h]
\includegraphics[width=0.49\columnwidth]{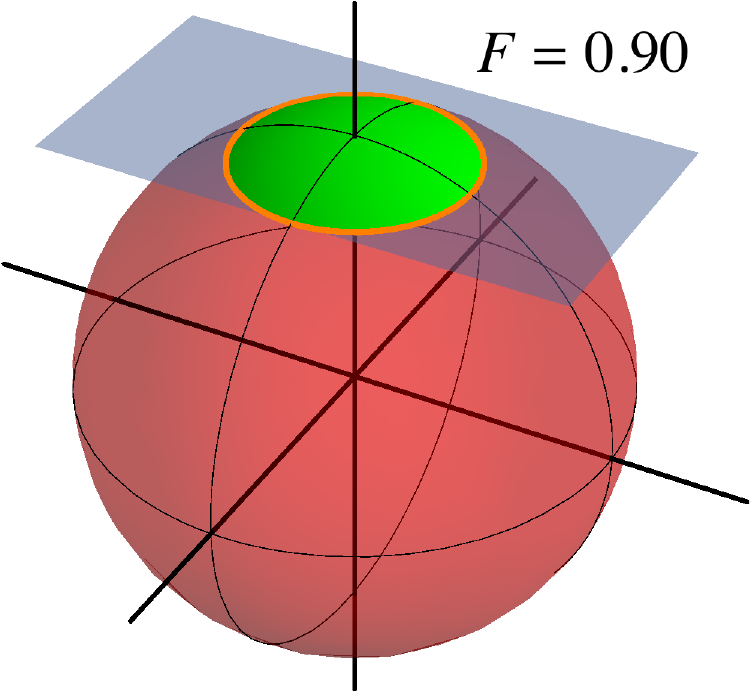}
\includegraphics[width=0.49\columnwidth]{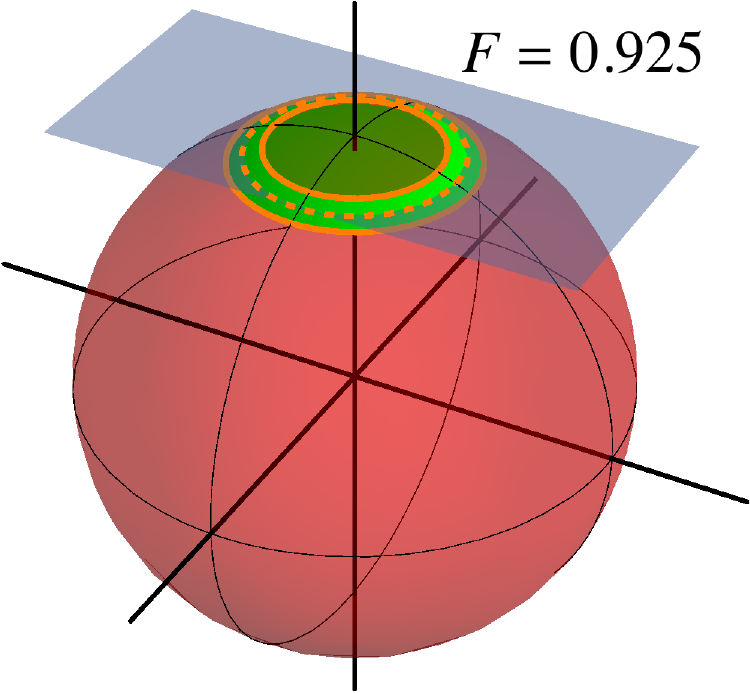}
\caption{(Color online) The green volumes represent single qubit states having
  fidelity larger than the threshold $F>0.9$ (left) or a fidelity
  $F=0.925\pm 0.025$ (right) to the target state
  $|0\rangle$.}\label{f:1q}
\end{figure}

In order to illustrate the point let us start with a very simple
example. Suppose you are given a qubit, aimed at being prepared
in the basis state $|0\rangle$, and guaranteed to have either a 
fidelity to the target state larger than a threshold, say $F>0.9$, 
or a given fidelity within a confidence interval, say $F=0.925\pm 0.025$. 
The situation is depicted in Fig. \ref{f:1q} where we show the 
corresponding  regions on the Bloch sphere.  \REVISION{As it is apparent from
the plots, neighboring states in terms of fidelity are compatible with a relatively 
large portion of the sphere that includes those states with different physical properties, 
e.g. the spin component in the $z$ direction}.
\par
The rest of the paper is devoted to illustrate few relevant, and ``more
dramatic'' examples, for two-qubit states and for continuous variable 
ones, where fidelity should be employed with caution to assess quantum 
resources. Indeed, our examples show that high values 
of fidelity may be achieved by pairs of states with considerably different 
physical properties, e.g. states containing quantum resources and states
of no value for quantum technology. Our examples are thus especially relevant 
for certification of quantumness in the presence of noise.
\par
The paper is structured as follows. In the next Section we address
two-qubit systems, focusing on both entanglement and discord of 
nearby Pauli diagonal states. The subsequent Sections are devoted 
to continuous variable systems: Section \ref{s:cv1} addresses 
certification of quantumness for single-mode squeezed thermal 
states and their displaced versions, whereas in Section \ref{s:cv2} we 
focus on entanglement and discord of two-mode squeezed thermal
states. Section \ref{s:out} closes the paper with some
concluding remarks.
\par
\section{Two-qubit systems}  
Let us consider the subset of {\em Pauli diagonal} (PD) two-qubit states
\begin{equation}\label{PDstate}
\rho=\frac{1}{4} \left ( \Id\otimes \Id+\sum_{j=1}^3c_j\sigma_j
\otimes\sigma_j \right )
\end{equation}
where $c_j$ are real constants, $\Id$ is the 
identity operator and 
$\sigma_j$ are Pauli matrices. The corresponding eigenvalues are
\begin{equation}\label{eigPD}\begin{split}
\lambda_0=\frac{1}{4} \left ( 1-c_1-c_2-c_3 \right )\\ 
\lambda_1=\frac{1}{4} \left ( 1-c_1+c_2+c_3 \right )\\ 
\lambda_2=\frac{1}{4} \left ( 1+c_1-c_2+c_3 \right )\\
\lambda_3=\frac{1}{4} \left ( 1+c_1+c_2-c_3 \right )
\end{split}\end{equation}
whose positivity implies constraints on coefficients $c_j$ for $\rho$ to describe
a physical state.
PD states in Eq. (\ref{PDstate}) have maximally 
mixed marginals (partial traces) $\rho^A=\rho^B=\Id/2$, $A$ and $B$ denoting 
the two subsystems. 
The choice of this subset stems from the fact that an analytic 
expression of the quantum discord is available \cite{Luo}, so we can 
compare quantum discord and entanglement of states within the PD class
for fixed values of fidelity.
The fidelity between two PD states may be expressed in terms 
of the eigenvalues in Eq. (\ref{eigPD}) as follows
\begin{equation}
{F} \left ( \rho_1,\rho_2 \right )=
\Big ( \sum_{k=0} ^3 \sqrt{\lambda_{k,1} \lambda_{k_2}} \Big )^2,
\end{equation} 
whereas entanglement, quantified by negativity, is given by 
\begin{equation}
{N}(\rho)=-2\sum_i\eta_i(\rho^{\tau_A}),
\end{equation} 
where $\eta_i(\rho^{\tau_A})$ are the negative eigenvalues 
of the  partial transpose $\rho^{\tau_A}$ with respect to the subsystem $A$ \cite{negativity}.
The quantum discord for PD states has been
evaluated in \cite{Luo}, and it is given by 
\begin{equation}
{D}(\rho)={I}(\rho)-\frac12 (1-c)\log_2(1-c)-\frac12 (1+c)\log_2(1+c)
\end{equation}
where ${I}(\rho)=2+\sum_{i=0}^3\lambda_i \log_2\lambda_i$ is the mutual 
information and the other terms are the result of the maximization of
the classical information. The quantity $c$ denotes the maximum 
$c\equiv\text{max}\{|c_1|,|c_2|,|c_3|\}$.
\begin{figure}[h]
\includegraphics[width=0.5\columnwidth]{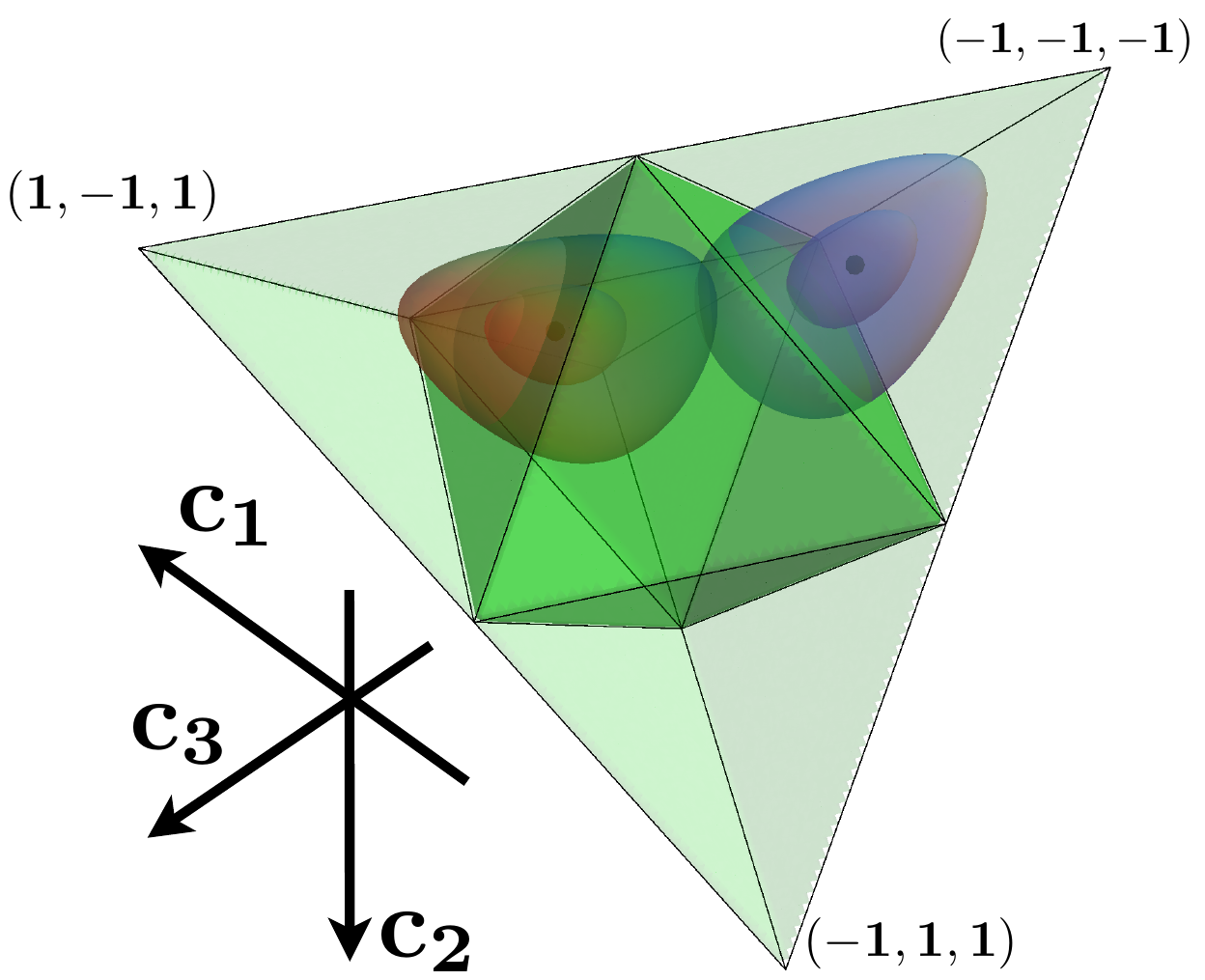}
\includegraphics[width=0.47\columnwidth]{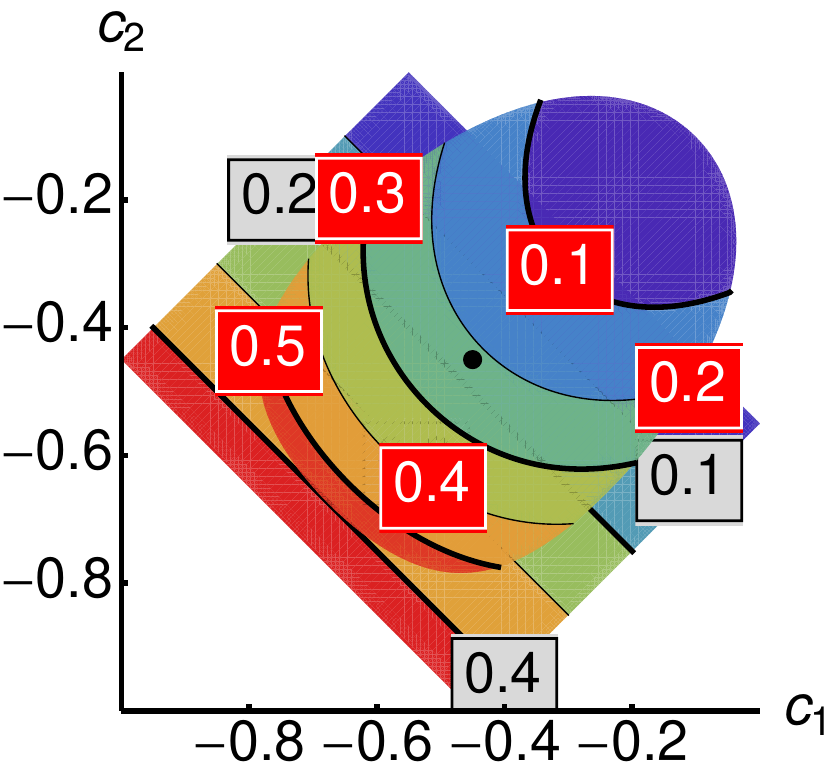}
\caption{\label{f:f2} (Color online) (Left panel): 
The tetrahedron represents the region of all physical PD states, whereas
the inner octahedron contains the separable ones. \REVISION{The balloons
centered in $c_1=c_2=c_3=-0.45$ (on the right of the panel) contain PD
states having fidelity ${F}>0.95$ and $F>0.99$ to the target Werner
(entangled) state.} The balloons on the left of the panel describe
states having fidelity $F>0.95$ and  $F>0.99$ to the separable PD state
with  $c_1=0.3$, $c_2=-0.3$, and $c_3=0.1$ .  (Right panel): the plot
describes PD states with fixed $c_3=-0.45$ and varying $\{ c1,c2\}$. We
show the ovoidal slice containing states having fidelity ${F}>0.95$ to
the target Werner state with $c_1=c_2=c_3=-0.45$ and the corresponding
rectangular region of entangled states.  Contour lines refer to
entanglement negativity (gray) and quantum discord (red).}
\end{figure}
\par 
Let us now consider a situation where the target state of, say, 
a preparation scheme, is a Werner state
$$\rho_W=\frac{1-c}{4} \Id\otimes \Id+c\ketbra{\Psi^-}{\Psi^-}\,,$$
i.e. a PD state with $c_1=c_2=c_3=-c$ and $c\in[0,1]$ and where 
$\ket{\Psi^-}=(\ket{01}-\ket{10})/\sqrt{2}$ is  one of the Bell states.
The Werner state $\rho_W$ is entangled for $c>\frac13$ and separable otherwise.
In particular, let us choose a target state with $c=0.45$ and address the
properties of PD states having fidelity larger than a threshold, say 
$F>0.95$ or $F>0.99$ to this target. Results are reported in the left 
panel of Fig. \ref{f:f2}, where 
the tetrahedral region is the region of physical 
two-qubit PD states and the separable states are confined to the inner octahedron.
The ovoidal regions (from now on the \emph{balloons}) contain the PD states with 
fidelity ${F}>0.95$ and $F>0.99$ to our target Werner state.
As it is apparent from the plot, both the balloons cross the separability 
border, thus showing that \REVISION{a ``high''
value of fidelity to the target should not be used as a benchmark for
creation of entanglement}, even assuming that the generated state
belongs to the class of PD states. The same phenomenon may lead one
to waste entanglement, i.e. to erroneously recognize an 
entangled state as separable on the basis of a high fidelity 
to a separable state, as it may happen to an initially maximally
entangled state driven towards the separability threshold by the
environmental noise.
As an example, we show in the left panel 
of Fig. \ref{f:f2} the balloons of states with fidelity $F>0.95$ 
and $F>0.99$ to a separable PD state with $c_1=0.3$, 
$c_2=-0.3$, and $c_3=0.1$.
\par
In the right panel of Fig. 
\ref{f:f2} we show 
the  ``slice'' of PD states with $c_3=-0.45$ and fidelity ${F}>0.95$ 
to the Werner target, together with the corresponding region of entangled 
states, and the contour lines of entanglement negativity and 
quantum discord. This plot clearly shows that high values 
of fidelity are compatible with large range of variation for both 
entanglement and discord.
\par
The fact that neighboring states may have quite different physical
properties has been recently investigated for quantum optical
polarization qubits \cite{edvd}. In particular, the discord of several
two-qubit states has been experimentally determined using partial and
full polarization tomography.  Despite the reconstructed states had
high fidelity to depolarized or phase-damped states, their discord has
been found to be largely different from the values predicted for these
classes of states, such that no reliable estimation procedure beyond
tomography may be effectively implemented, and thus questioning the
use of fidelity as a figure of merit to assess quantum correlations.
Indeed, when full tomography is performed, fidelity is used only 
to summarize the overall quality of the reconstruction  
\cite{anto1,anto2,nat1,nat2} and thus correctly convey also 
the information obtained about quantum resources.
\par
\section{Single-mode Gaussian States} 
\label{s:cv1}
Here we address the use of fidelity to assess quantumness of single-mode
CV states. In particular, in Section \ref{s:cv1a} we address nonclassicality 
of squeezed thermal states, whereas Section \ref{s:cv1b} is devoted the 
subPossonian character of their displaced version.
\subsection{Squeezed thermal states}
\label{s:cv1a}
\begin{figure}[h!]
\includegraphics[width=0.49\columnwidth]{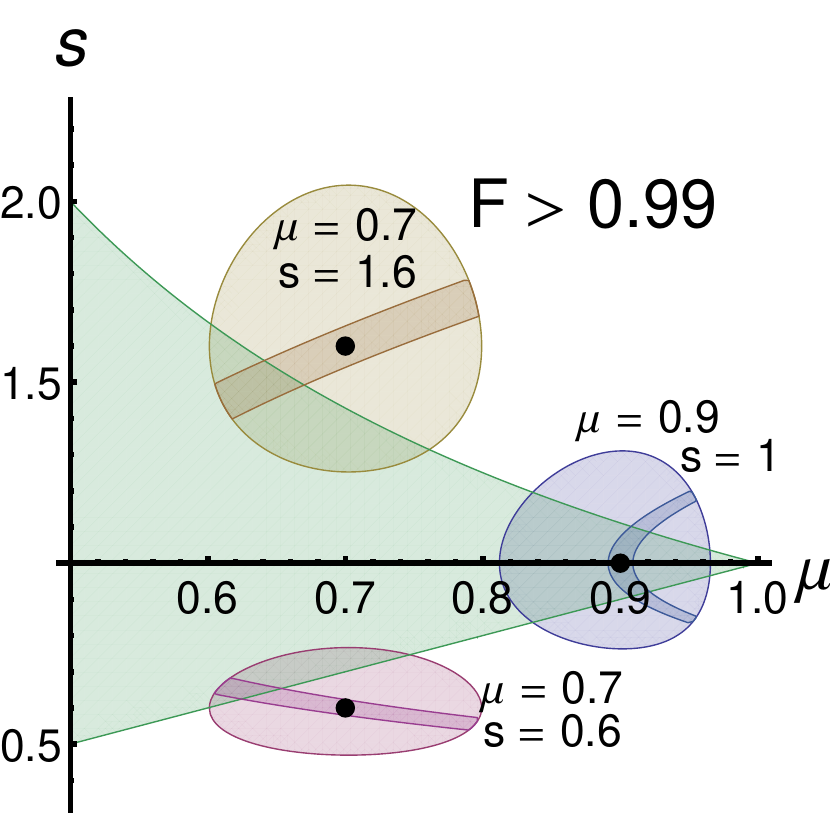}
\includegraphics[width=0.49\columnwidth]{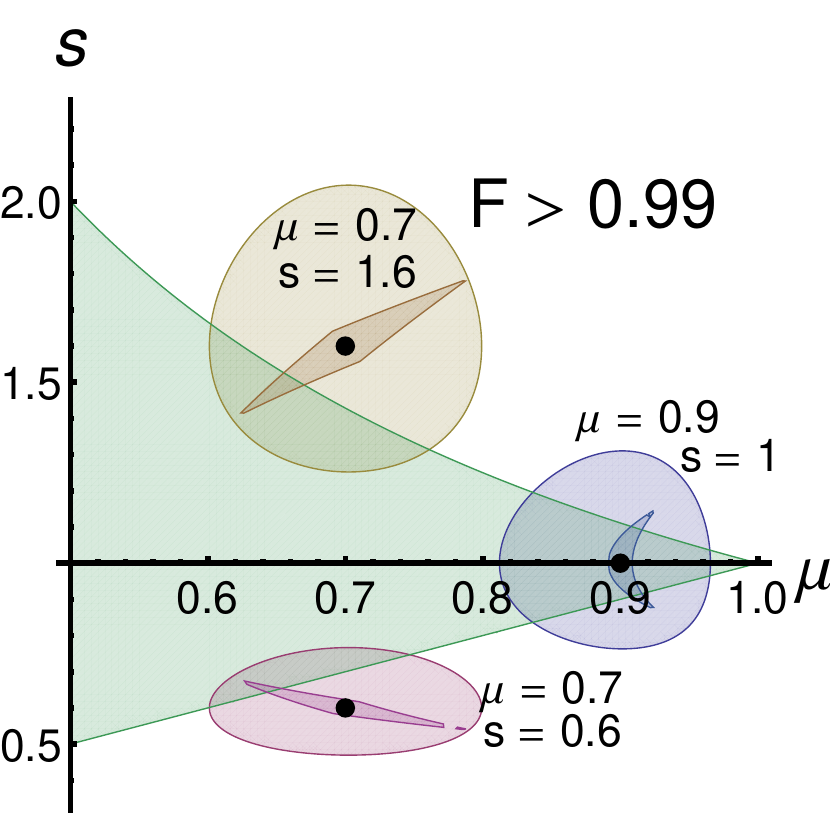}
\caption{\label{f:f3} (Color online) The plots show the 
region of classicality (the triangular-like green regions) together
with the balloons of $\stso$ having fidelity larger than $F_{s\mu}>0.99$ to 
three $\stso$ chosen as targets: a classical thermal state with 
$s=1$ and $\mu=0.9$ and two nonclassical $\stso$ with 
$\mu=0.7$ and $s=0.6$ and $s=1.6$ respectively. In the left panel
the stripes of states close to the targets contain states
having $F_{s\mu}>0.99$ and mean photon numbers which differ at most $10\%$
from that of the target. In the right panel the states close
to the targets satisfy the additional constraints of having number
fluctuations within a $10\%$ interval from that of the targets.}
\end{figure}

Let us now consider single-mode CV
systems and start with Gaussian state preparations of the form 
\begin{equation}
\rho_{s\mu}= S(r)\nu (N)S^\dag(r)
\end{equation}
i.e. single-mode squeezed
thermal states ($\stso$) with real squeezing, $S(r)=\exp \{ \frac12 r (a^{\dag 2} -
a^2) \} $ and  $N$ thermal photons, $\nu (N)= N^{a^\dag a}/(1+N)^{a^\dag
a+1}$. This class of states have zero mean and covariance matrix 
(CM) given by 
\begin{equation}
\sigma=\frac{1}{2\mu}
\begin{pmatrix}
1/s& 0\\
0 & s
\end{pmatrix},
\end{equation}
where $\mu=(2
\sqrt{\det{\sigma}})^{-1} = (2N+1)^{-1}$ is the purity of $\rho_{s\mu}$ and 
$s=e^{-2r}$ is the squeezing factor. $\stso$ are
nonclassical, i.e. they show a singular Glauber P-function, when
$s<\mu$ or $s>1/\mu$ \cite{CTLee}. 
Fidelity between two $\stso$
is given by \cite{twa96,scu98} 
\begin{equation}
F_{s\mu}=\frac{1}{\sqrt{\Delta + \delta} - \sqrt{\delta}}
\end{equation}
where 
$$\Delta=\det[\sigma_1 + \sigma_2] \qquad 
\delta=4\prod_{k=1}^2 \left[\det[\sigma_k]-\frac{1}{4}\right]\,,$$ 
$\sigma_1$ and
$\sigma_2$ being the CM of the two states.
In Fig. \ref{f:f3} we report the region of classicality together
with the balloons of $\stso$ having fidelity larger than $F_{s\mu}>0.99$ to 
three $\stso$ chosen as targets (one classical thermal
state and two nonclassical thermal squeezed states).
\par
\REVISION{As it is apparent from the plot, the balloons have large overlaps
with both the classical and the nonclassical region, such that 
fidelity cannot be used, for this class of states, to certify the creation 
of quantum resources.} This feature is only partially cured 
by imposing additional constraints to the set of states under
examination \cite{dodonov}. As for example, in the left panel of Fig. \ref{f:f3}
we show the ``stripes'' of states that have both a fidelity $F_{s\mu}>0.99$ 
{\em and} a mean photon numbers $\langle n\rangle$, 
(i.e. the mean energy of state) 
which differ at most $10\%$ from that of the target.  In the right 
panel we show the regions of states satisfying also 
the additional constraints of having photon number
fluctuations $\langle \Delta n^2\rangle$ 
within a $10\%$ interval from that of the targets.
Overall, we have strong evidence that fidelity should not be used to
certify the presence of quantumness, and that this behavior persists
even when we add quite stringent constraints to delimit the class
of states under investigations. In fact, only by performing the 
full tomographic reconstruction of the state one imposes a suitable 
set of constraints to make fidelity a fully meaningful figure of merit 
\cite{Jar09}. In this case, as already mentioned for qubits, fidelity 
represents a summary of the precision achieved by the full 
tomographic reconstruction.
\par
\subsection{Displaced squeezed thermal states}
\label{s:cv1b}
\begin{figure}[t]
\includegraphics[width=.49\columnwidth]{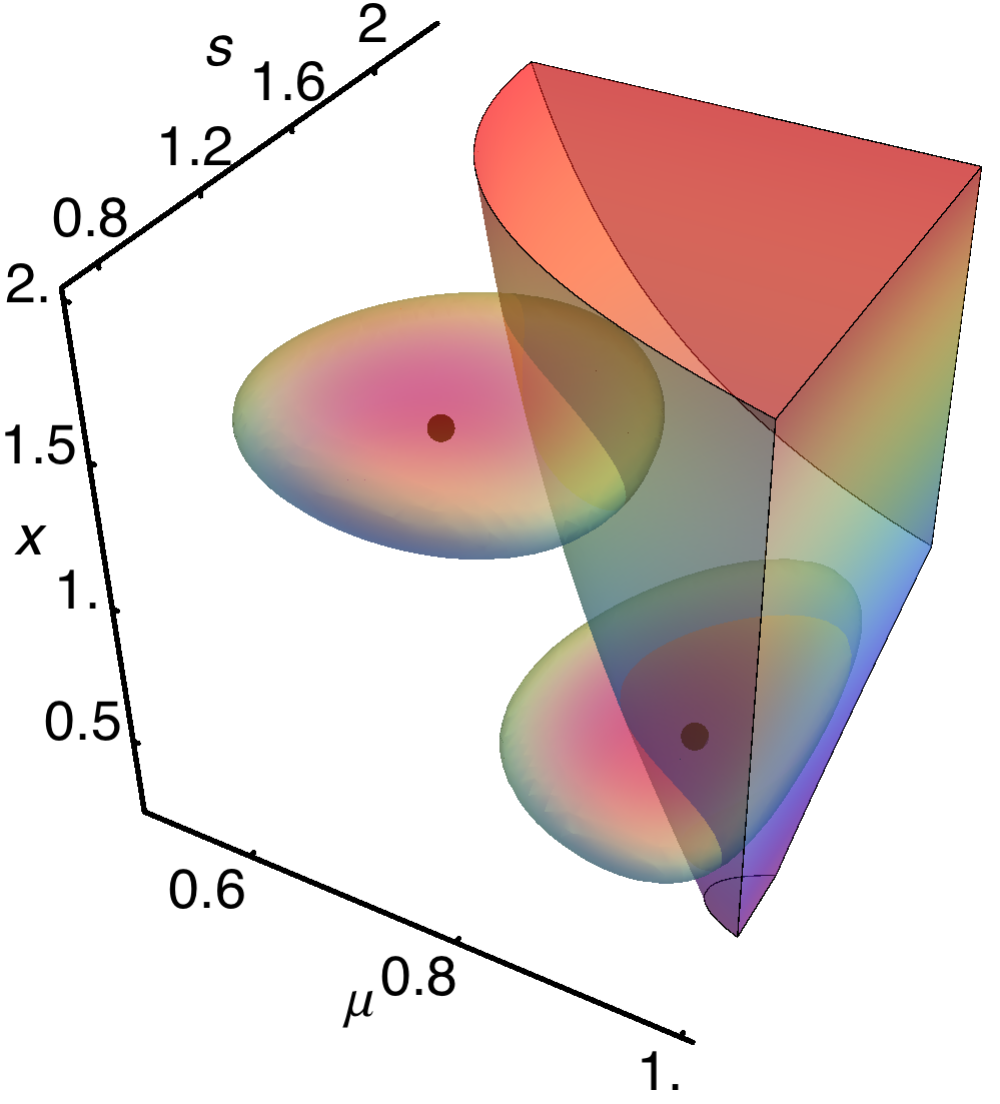}
\includegraphics[width=.49\columnwidth]{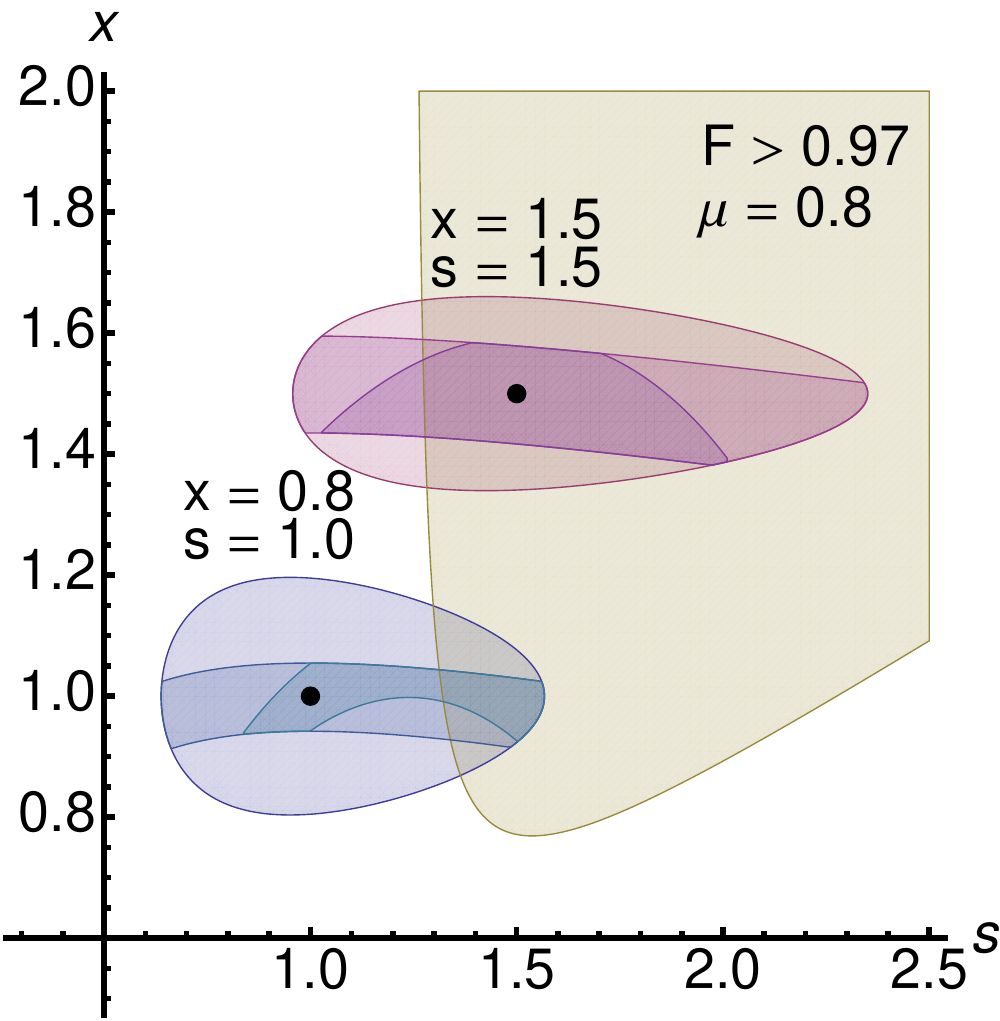}
\caption{\label{f:f4}(Color online) (Left): subPoissonian region for
$\rho_G$ states as a function of the purity $\mu$, the squeezing $s$,
and the displacement $x$, together with the balloons of states having
fidelity larger than $F_G>0.97$ to a nonclassical target with $\mu=0.9$,
$s=1.4$, and $x=0.5$ and a classical one with $\mu=0.7$, $s=1.2$, and
$x=1.5$. (Right): The subPoissonian region for a fixed value of
purity $\mu=0.8$ as a function of squeezing and displacement, together with
the balloons of states having fidelity larger than $F_G>0.97$ to the
target states having $x=1.5$ and $s=1.5$ (subPoissonian) or $x=0.8$ and
$s=1.0$ (superPoissonian). We also show the subregions of states having
mean photon number and number fluctuations which differ at most $10\%$
from those of the target.} \end{figure}

When only intensity measurements may be performed, 
nonclassicality of a single-mode state may be assessed by 
the Fano Factor \cite{HP82}, which is defined as the ratio of the 
photon number fluctuations over the mean photon number
$R= \langle \Delta n^2 \rangle/\langle n
\rangle$. One has $R=1$ for coherent states, while a smaller 
value is a signature of nonclassicality since sub-Poissonian statistics 
cannot be described in classical terms.
In order to illustrate the possible drawbacks of fidelity in certifying 
this form of quantumness, let us
consider displaced version of $\stso$
\begin{equation}
\rho_G=D(x)\rho_{s\mu}D^\dag(x),
\end{equation}
where $D(\alpha)=\exp\{\alpha a^\dag
- \bar\alpha a\}$ is the displacement operator and we chose 
real displacement $\alpha=x\in {\mathbbm R}$. The CM is determined
by $\rho_{s\mu}$ whereas the displacement change only the mean values
of the canonical operators.  The fidelity between two Gaussian states 
of the form $\rho_G$ is given by \cite{scu98} 
\begin{equation}
F_G = 
\exp\{-(\mathbf X_1 - \mathbf X_2)^T(\sigma_1+\sigma_2)^{-1}
(\mathbf X_1 - \mathbf X_2)\} F_{s\mu}
\end{equation}
where $\mathbf X=(x,0)$.
In the left panel of Fig. \ref{f:f4} we show the region of
sub-Poissonianity as a function of the purity, the squeezing factor, and 
the displacement of states $\rho_G$. We also show the balloons of 
states with fidelity larger than $F_G>0.97$ to two $\rho_G$ target
states: a subPoissonian state corresponding to $\mu=0.9$, $s=1.4$, and 
$x=0.5$ and a superPoissonian one with $\mu=0.7$, $s=1.2$, and 
$x=1.5$. Despite the high value of fidelity (notice that fidelity
decreases exponentially with the displacement amplitude) both 
the balloons crosses the Poissonian border, and the parameters of the 
states that may differ considerably from the targeted
ones. In the right panel of Fig. \ref{f:f4} we show the subPoissonian region 
for a fixed value of purity $\mu=0.8$ as a function of squeezing and 
displacement, together with the balloons of states having fidelity 
larger than $F_G>0.97$ to a pair of target states: a subPoissonian 
state with parameters $x=1.5$ and $s=1.5$ and a superPoissonian one with 
$x=0.8$ and $s=1.0$. We also show the subregions of states having 
mean photon number and number fluctuations which differ at most $10\%$ 
from those of the target. We notice that even restricting attention
to states with comparable energy and fluctuations, fidelity is
not able to discriminate states having quantum resources or not.
\par
\section{Two-mode Gaussian States}
\label{s:cv2}
Here we focus on a relevant subclass 
of two-mode Gaussian states: the so-called two-mode squeezed 
thermal states ($\stst$) described by density operators of the form
\begin{equation}
\rho_{N\beta\gamma} = S_2(r)\nu(n_1)\otimes\nu(n_2)S_2^\dag(r)
\end{equation}
where $S_2(r)=\exp\{r(a^\dag b^\dag-a b)\}$ is the two-mode squeezing 
operator with \REVISION{real parameter $r$ and $\nu(n_k)$, $k=1,2$ are thermal states 
with $n_k$ photon number on average}. The class of states
$\rho_{N\beta\gamma}$ is fully described by three 
parameters: the total mean photon number 
$N$, the two-mode squeezing fraction
$\beta$ and the single-mode fraction of thermal photons:
$\gamma$
\begin{equation}\label{parameter}\begin{split}
&N=\langle a^\dag a + b^\dag b\rangle\\
&\beta=\frac{2 \sinh^2 r}{N}\\
&\gamma=\frac{n_1}{n_1+n_2}.
\end{split}\end{equation}
The CM of $\stst$ may be written in the block form
\begin{equation}
\sigma=\frac{1}{2}\begin{pmatrix}
A\, \mathbb{I}& C\, \sigma_z\\
C\, \sigma_z & B\, \mathbb{I}
\end{pmatrix}
\end{equation}
with the coefficients parametrized according to (\ref{parameter}):
\begin{equation}\begin{split}
A&=1+\frac{2\gamma(1-\beta)N+\beta N (1+N)}{1+\beta N}\\
B&=1+\frac{2(1-\gamma)(1-\beta)N+\beta N (1+N)}{1+\beta N}\\
C&=\frac{(1+N)\sqrt{\beta N(2+\beta N)}}{1+\beta N}.
\end{split}\end{equation}
A squeezed thermal state is
separable iff $\tilde d_- \geq \frac12$, 
where $
\sqrt{2}\tilde d_\pm 
= \sqrt{A^2+B^2+2C^2\pm (A+B)\sqrt{(A-B)^2+4C^2}}$ are the 
symplectic eigenvalues. Gaussian B-discord, i.e. the difference 
between the mutual information and the maximum amount of 
classical information obtainable by {\em local Gaussian} 
measurements on system B, may be analytically evaluated for 
$\stst$ \cite{gd}, leading to 
\begin{equation}
D(\rho_{N\beta\gamma})=h(B) 
- h(d_-)-h(d_+) + h\left ( \frac{A-C^2}{B+\frac12} \right )
\end{equation}
where $h(x)=
(x+\frac12) \ln (x+\frac12) 
-(x-\frac12) \ln(x+\frac12)$.
Finally, fidelity between two $\stst$ is given 
by \cite{sorin,Marian,Oli12}
\begin{equation}
{F}_{N\beta\gamma} = 
\frac{(\sqrt{{X}}+\sqrt{{X}-1})^2}{\sqrt{\det[\sigma_1+\sigma_2]}}
\end{equation}
where 
\begin{align}
X&=2\sqrt{{E_1}}+2\sqrt{{E_2}}+\frac{1}{2}\,,\notag\\ 
E_1&=\frac{\det[\Omega\,\sigma_1\,\Omega\,\sigma_2]-
\frac14}{\det[\sigma_1+\sigma_2]}\,,\notag\\
E_2&=\frac{\det[\sigma_1+\frac{\imm}{2}\Omega]\det[\sigma_2+\frac{\imm}{2}\Omega]}{
\det[\sigma_1+\sigma_2]}\,,\notag
\end{align}
$\Omega$ being the $2$-mode symplectic form \cite{Oli12}
\begin{equation}
\Omega=\omega \oplus \omega \qquad \omega=
\left(\begin{array}{cc}0&1\\ -1&0
\end{array}\right)\,. \notag
\end{equation}
\par
\begin{figure}[h!]
\includegraphics[width=0.49\columnwidth]{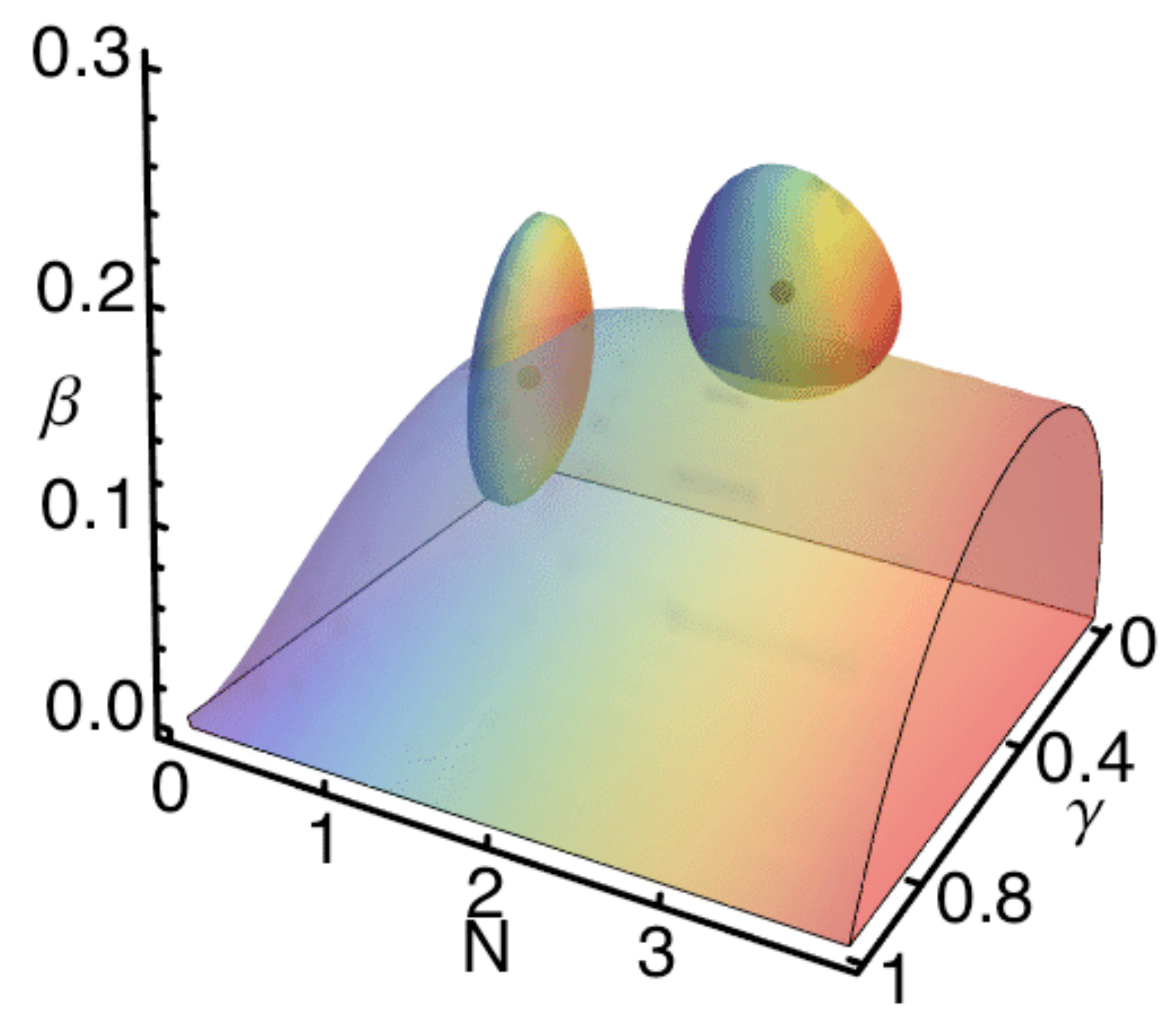}
\includegraphics[width=0.49\columnwidth]{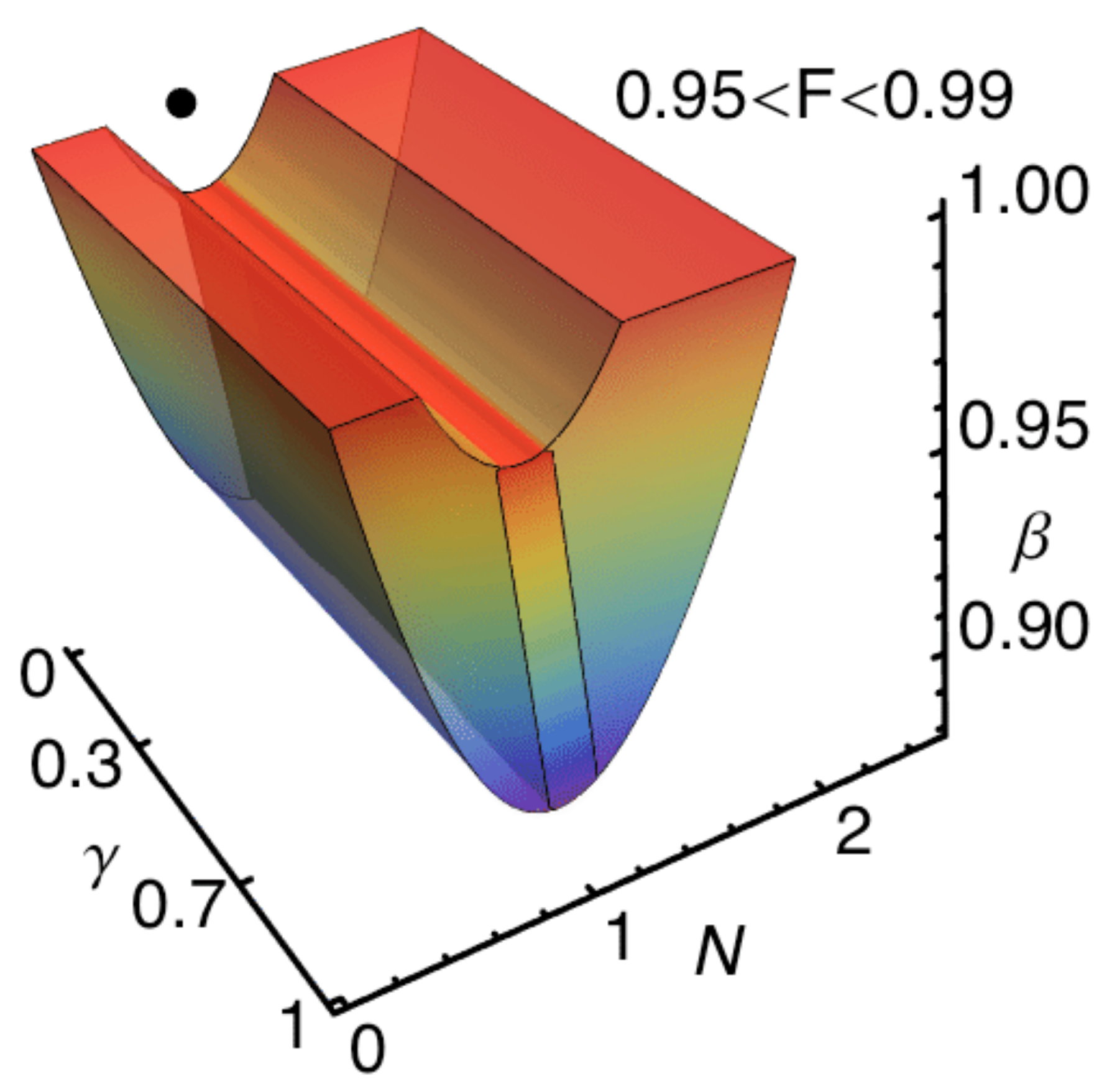}
\caption{\label{f:f5}(Color online) (Left): Separability region of $\stst$ in terms 
of the three parameters $N$, $\beta$ and $\gamma$ together
with the balloons of states having ${F}_{N\beta\gamma}>0.99$ with two 
target states: an entangled $\stst$ with parameters 
$N=2.5$, $\beta=0.2$, $\gamma=0.5$ and a separable one
with $N=1$, $\beta=0.13$ and $\gamma=0.5$. (Right): the region of 
of states having a fidelity in the range
$0.95 < F_{N\beta\gamma}<0.99$ to a two-mode squeezed vacuum, 
$N=1$ and $\beta=1$. We also show the stripe of states having 
a mean photon number in the range $0.9<N<1.1$.}
\end{figure}

In the left panel of Fig. \ref{f:f5} we show the separability
region in terms of the three parameters $N$, $\beta$ and $\gamma$ together
with the balloons of states having ${F}_{N\beta\gamma}>0.99$ with two 
target states: an entangled $\stst$ with parameters 
$N=2.5$, $\beta=0.2$, $\gamma=0.5$ and a separable one
with $N=1$, $\beta=0.13$ and $\gamma=0.5$. As it is apparent from the
plot, both balloons cross the separability border and have a
considerable overlap to both regions, thus making fidelity of little use 
to assess entanglement in these kind of systems. 
\par
\begin{figure}[h!]
\includegraphics[width=0.49\columnwidth]{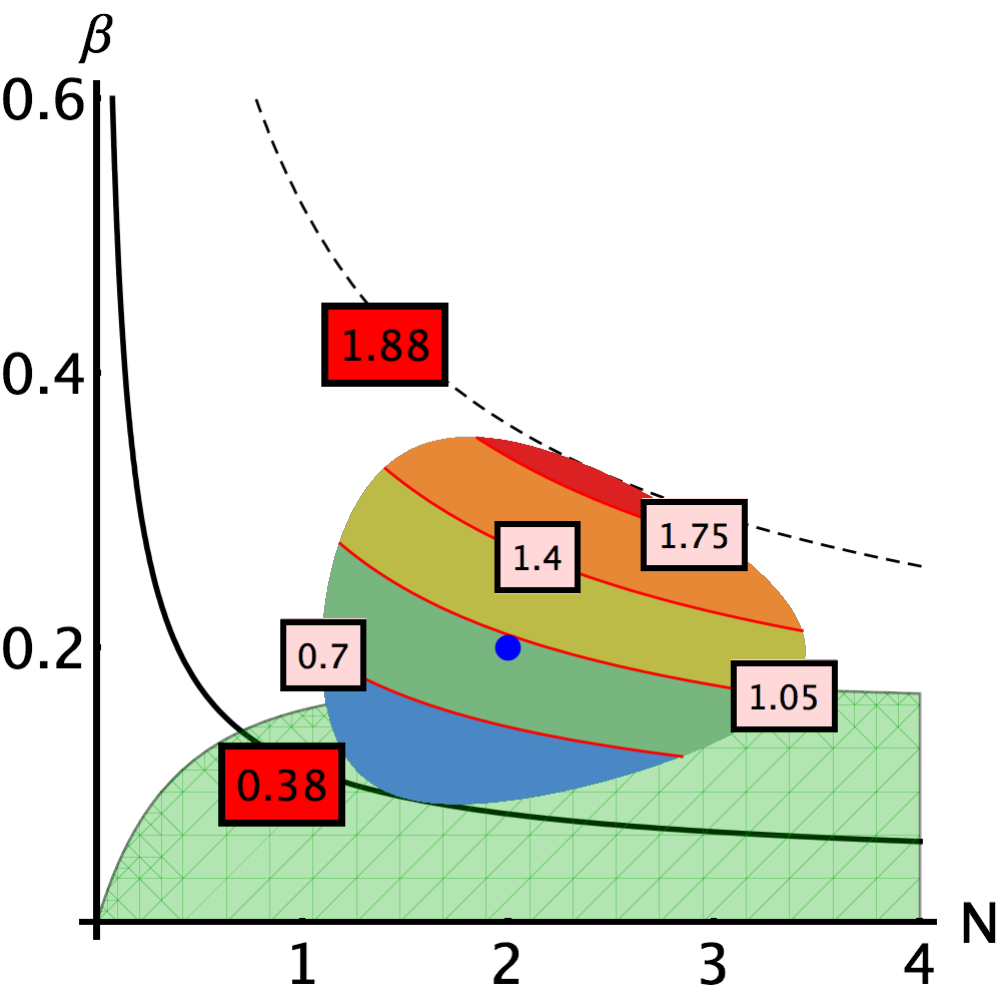}
\includegraphics[width=0.49\columnwidth]{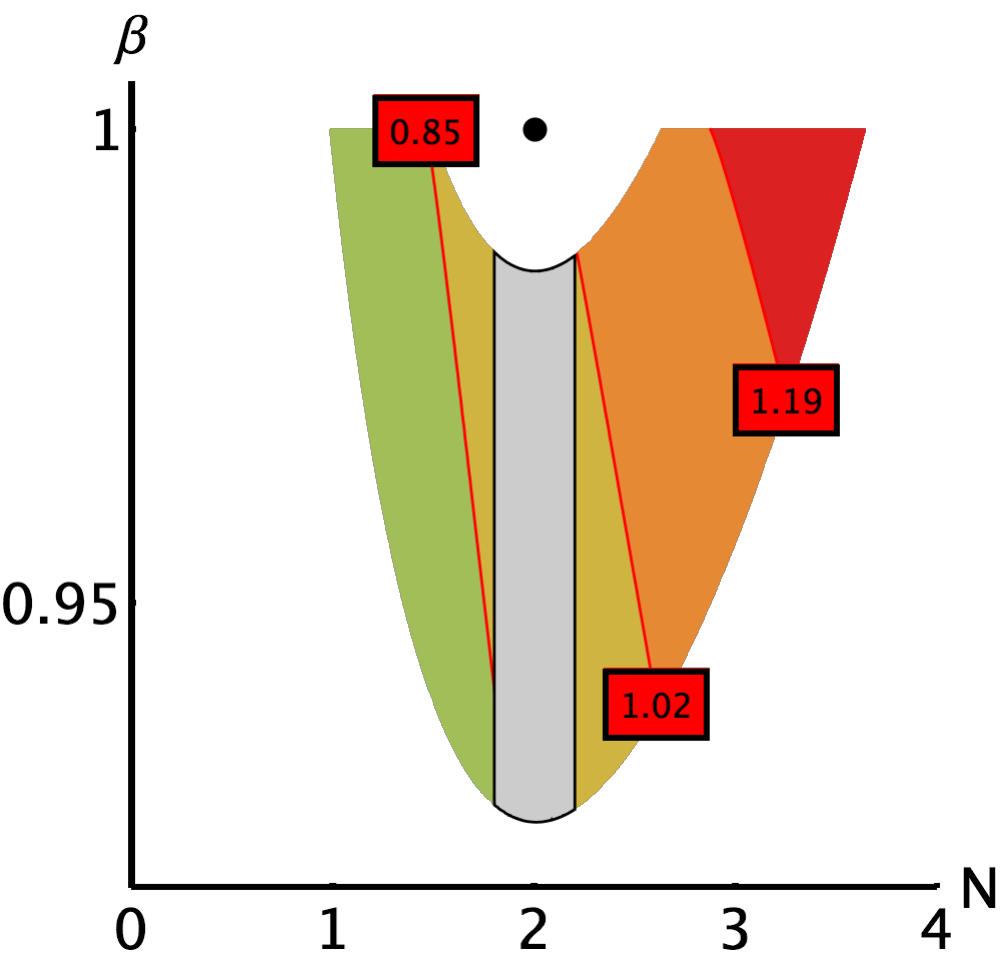}
\caption{\label{f:f6}(Color online) (Left): Contour lines of Gaussian B-discord in the
region of $\text{STS}_2$ having fidelity $F_{N\beta\gamma}>0.95$ to an
entangled target state with $N=2,\beta=0.2$ and $\gamma=0.5$. The relative
discord, rescaled to that of the target state
($D(\rho_{2,0.2,0.5})=0.22$), ranges from $0.38$ to $1.88$.  (Right):
Variations of the relative Gaussian B-discord in a region of
$\text{STS}_2$ with fidelity $0.95 < F_{N\beta\gamma} < 0.99$ to a
two-mode squeezed vacuum state ($N=2$ and $\beta=1$). In evidence the
constrained region of states having the $10\%$ of energy fluctuations
around $N=2$.} 
\end{figure}

Another phenomenon arising from 
benchmarking with fidelity is illustrated in the
right panel of Fig. \ref{f:f5}, where we report the region of 
 states having a fidelity in the range
$0.95 < F_{N\beta\gamma}<0.99$ to a {\em two-mode squeezed vacuum}, 
i.e. a maximally entangled state with $N=1$ and $\beta=1$. The
emphasized sector corresponds to states that also have a mean photon
number not differing more than $10\%$ from the target, i.e. in 
the range $0.9<N<1.1$. As a matter
of fact, the total photon number $N$ and the squeezing 
fraction $\beta$ in this region may be 
considerably different from the targeted one and, in addition, the
states with comparable energy are the least entangled in the region.
Finally, in Fig. \ref{f:f6} we show the range of variation of
Gaussian B-discord compatible with high values of fidelity. In the
left panel we consider a non-separable target state with discord
$D(\rho_{2,0.2,0.5})=0.22$ and a region of $\text{STS}_2$ states with
fidelity $F_{N\beta\gamma}>0.95$. The region of separability (green) is
crossed by a non negligible set of states and the relative variations of
the discord is considerably large, ranging from $0.38$ to $1.88$. In the
right panel of Fig. \ref{f:f6} we show again the wide range of variation
of Gaussian B-discord for a set of $\text{STS}_2$ states with fidelity
$0.95<F(\rho_{N\beta\gamma})<0.99$ to a target two-mode squeezed vacuum
state with $N=2$. The high discrepancy in the relative discord can be
only partially limited by constraining the mean photon number $N$ with
fluctuations of the $10\%$.  
Notice that also in the case of two modes, full Gaussian tomography 
\cite{FullCM,bla12}
is imposing a suitable set of constraints to make fidelity a meaningful
figure of merit to summarize the overall quality of the reconstruction. 
\par
\par
\section{Conclusions}
\label{s:out}
In conclusion, we have shown by examples that being close in the Hilbert 
space may not imply being close in terms of quantum resources. In particular,
we have provided quantitative examples for qubits and
CV systems showing that pairs of states with high fidelity may include separable and
entangled states, classical and nonclassical ones, and states with very
different values of quantum of Gaussian discord. \par 
Our results 
make apparent that in view of its wide use in quantum technology, fidelity is a
quantity that should be employed with caution to assess quantum
resources. In some cases it may be used in conjunction with additional
constraints, whereas in the general
situation it should be mostly used as an overall figure of merit, summarizing
the findings of a full tomographic reconstruction.
\begin{acknowledgments}
This work has been supported by the MIUR project FIRB-LiCHIS-
RBFR10YQ3H. MGAP thanks Claudia Benedetti 
for useful discussions. 
\end{acknowledgments}

\end{document}